\documentclass[11pt]{article}
\usepackage[total={6in, 9.4in}]{geometry}
\usepackage{url}

\usepackage{caption}

\usepackage{amsthm}
\theoremstyle{definition}

\usepackage{afterpage}

\usepackage{amsmath}
\usepackage[nameinlink, capitalise]{cleveref}

\usepackage[english]{babel}    
\usepackage[strings]{underscore}   

\usepackage{booktabs} 

\usepackage{graphicx}

\begin{document}
\title{On the Estimation of the Number of Unreachable Peers in the Bitcoin P2P Network by Observation of Peer Announcements
}
\author{Matthias Grundmann
		\and Hedwig Amberg
		\and Hannes Hartenstein
}
\date{%
    Institute of Information Security and Dependability (KASTEL)\\%
    Karlsruhe Institute of Technology (KIT), Karlsruhe, Germany%

    \{matthias.grundmann,hannes.hartenstein\}@kit.edu
}

\maketitle              

\makeatletter
\newcommand{\addrIPs}[1]{A_{#1}}
\newcommand{\publicIPs}[1]{P_{#1}}
\newcommand{\estimateUnreachable}[1]{U_{#1}}
\newcommand{\quickCount}[2]{q^{#1}_{#2}}
\newcommand{\slowCount}[2]{s^{#1}_{#2}}
\newcommand{\methodName}{PAL}

\makeatother

\begin{abstract}

Bitcoin is based on a P2P network that is used to propagate transactions and blocks of the blockchain.
While the P2P network design intends to hide the topology of the P2P network to impede adversarial actions against peers, information about the topology is required to understand the network from a scientific point of view, to build realistic models used for simulations, and to use these models to optimize protocols of the P2P networks. Thus, there is a natural tension between the
`desire' for unobservability on the one hand, and for observability on the other hand. On a middle ground, one would at least be interested on some statistical features of the Bitcoin network like the number of peers that participate in the propagation of transactions and blocks.
This number is composed of the number of reachable peers that accept incoming connections and unreachable peers that do not accept incoming connections.
Despite not accepting incoming connections, unreachable peers open several outgoing connections to other peers and participate in the propagation of transactions and blocks.
While the number of reachable peers can be measured, it is inherently difficult to determine the number of unreachable peers, exactly because one cannot connect to them.
Thus, the number of unreachable peers can only be estimated based on some indicators.
In this paper, we first define our understanding of unreachable peers and then propose the \methodName{} (Passive Announcement Listening) method which gives an estimate of the number of unreachable peers by observing \textsc{addr} messages that announce active IP addresses in the network.
The \methodName{} method allows for detecting unreachable peers that use flags to indicate that they provide services useful to the P2P network.
In conjunction with previous methods, the \methodName{} method can help to get a better estimate of the number of unreachable peers. We use the \methodName{} method to analyze data from a long-term measurement of the Bitcoin P2P network that gives insights into the development of the number of unreachable peers over more than five years from 2015 to 2020.
Results show that about 31,000 unreachable peers providing useful services were active per day at the end of the year 2020. An empirical validation indicates that the approach finds about 50\,\% of the unreachable peers that provide useful services.

\end{abstract}


\section{Introduction}

In Bitcoin \cite{nakamoto_bitcoin:_2008}, transactions and blocks are propagated by peers that are connected in a peer-to-peer (P2P) network.
The protocol used for the propagation of transactions and blocks affects the whole system's performance security:
The propagation delay of blocks is connected to the number of forks, i.e. temporal inconsistencies in the blockchain \cite{decker_information_2013}.
The protocol's bandwidth consumption limits the maximum number of connections \cite{naumenko_erlay_2019-1} which limits the resilience against eclipse attacks \cite{heilman_eclipse_2015}.
The protocol can also leak information about the creator of a transaction \cite{fanti_deanonymization_2017} and partial information about the topology of the network \cite{neudecker_timing_2016,grundmann_exploiting_2019,delgado-segura_txprobe_2019}.
Analyzing the protocol with regard to these aspects and validating proposals for improvements can hardly be done in  the live P2P network.
Thus, researchers and developers use simulations of the P2P network for these tasks \cite{neudecker_security_2019,naumenko_erlay_2019-1}.
Such simulations require a model of the topology of the P2P network.
However, the real topology of the Bitcoin P2P network is unknown to the public\footnote{Approaches to gain (partial) information about the topology have been proposed \cite{biryukov_deanonymisation_2014,miller_discovering_2015,neudecker_timing_2016,grundmann_exploiting_2019,delgado-segura_txprobe_2019}. However, they either do not work anymore or require a strong attacker to get the complete P2P network's topology.} because the protocol is designed to hide information about the topology that could support adversarial actions against peers \cite{fanti_deanonymization_2017,heilman_eclipse_2015}.
Therefore, simulations model the topology based on certain characteristics that can be measured or estimated without having a method for the measurement built into the protocol itself.
One of these characteristics and the object of this paper is the number of peers that participate in the propagation of transactions and blocks.

To form the P2P network, each peer creates by default 10 outgoing connections to other peers.
A peer is a running instance of a Bitcoin software that is connected to at least one other instance of a Bitcoin software.
Peers are encouraged to connect to multiple peers to reduce chances of being victim of an eclipse attack \cite{heilman_eclipse_2015}.
Not every peer accepts incoming connections, e.g., because a peer is behind a NAT or a firewall or a peer is configured to block incoming connections.
Thus, the peers can be categorized into two groups: reachable peers that accept incoming connections and unreachable peers that do not accept incoming connections.
Despite the name, unreachable peers play an active role in the P2P network.
While unreachable peers do not accept incoming connections, they open several outgoing connections to other peers and participate in the propagation of transactions and blocks just as reachable peers do.
Hence, for the propagation of transactions and blocks both, the reachable and unreachable peers are relevant.
Franzoni and Daza \cite{franzoni_improving_2020} recently presented how the robustness and efficiency of the P2P network can be improved by giving unreachable peers a special role in the propagation of transactions.

There exist projects\footnote{E.g., \url{https://bitnodes.io/} and \url{https://dsn.kastel.kit.edu/bitcoin/}\label{footnote-projects}} continuously measuring the number of reachable peers.
However, unreachable peers are mostly invisible because one cannot connect to them.
Although we can {\it define} the set of unreachable peers, we cannot {\it retrieve} it using measurements: We can only {\it estimate} the number of unreachable peers.
There are two ways to get such an estimate:
The first way is to observe a fraction of unreachable peers and extrapolate the whole number of unreachable peers.
This can be done by running a reachable peer that accepts incoming connections from unreachable peers.
The second way is to observe effects that are caused by unreachable peers and infer their number from these observations.
In this paper, we present an approach that uses the second way to estimate the number of unreachable peers.
This approach, that we call the \methodName{} (Passive Announcement Listening) method, relies on observing address announcements that are forwarded by reachable peers in the network.
The approach is only able to count unreachable peers that offer services to the network such as storing the latest 288 blocks of the blockchain.
Addresses of unreachable peers that do not announce such useful services will not be forwarded by other peers and thus not be counted.
As there is no ground truth available, we validate our approach by verifying our assumptions and by comparing the results of our approach to an observation of a fraction of unreachable peers.
We show that the approach detects most reachable peers and about 50\,\% of unreachable peers that provide useful services.
Previous work has estimated the number of unreachable peers to be around 16,000 peers \cite{neudecker_timing_2016}, 54,000 peers \cite{naumenko_erlay_2019-1}, 100,000 peers \cite{biryukov_deanonymisation_2014}, and 155,000 peers \cite{wang_towards_2017}.
The wide range of differences comes not only from different measuring times and methods but also from the fact that different definitions of the term \textit{unreachable peer} are used.

Thus, we start by defining the term \textit{unreachable peers} and other relevant terms in the following section.
An overview of related work will be given in \cref{sec-related-work}.
We describe the Bitcoin protocol and the behavior of the most common Bitcoin implementation in \cref{sec-background}.
Then, we present the \methodName{} method in \cref{sec-method} and present the results of applying the method to data collected from the Bitcoin P2P network between 2015 and 2020.
We validate the method in \cref{sec-validation} and conclude in \cref{sec-conclusion}.

\section{Definitions and Problem Statement}
\label{sec-problem-statement}
We refer to an implementation of the Bitcoin protocol as Bitcoin software.
As stated above, we define a \textit{peer} as a running instance of a Bitcoin software that is connected to at least one other running instance of a Bitcoin software.
We expect most peers, however, to be connected to multiple peers in order to reduce chances of being a victim of an eclipse attack.
A Bitcoin P2P network consists of peers that are directly or indirectly connected to each other.
Because this definition allows multiple Bitcoin P2P networks, we refer to \textit{the} Bitcoin P2P network as the Bitcoin P2P network that includes the peers that mine blocks with more computation power than the peers in every other Bitcoin P2P network.
In the following, we consider only peers that are part of \textit{the} Bitcoin P2P network.

\begin{figure}[tbp]
	\centering
	\includegraphics[width=12cm]{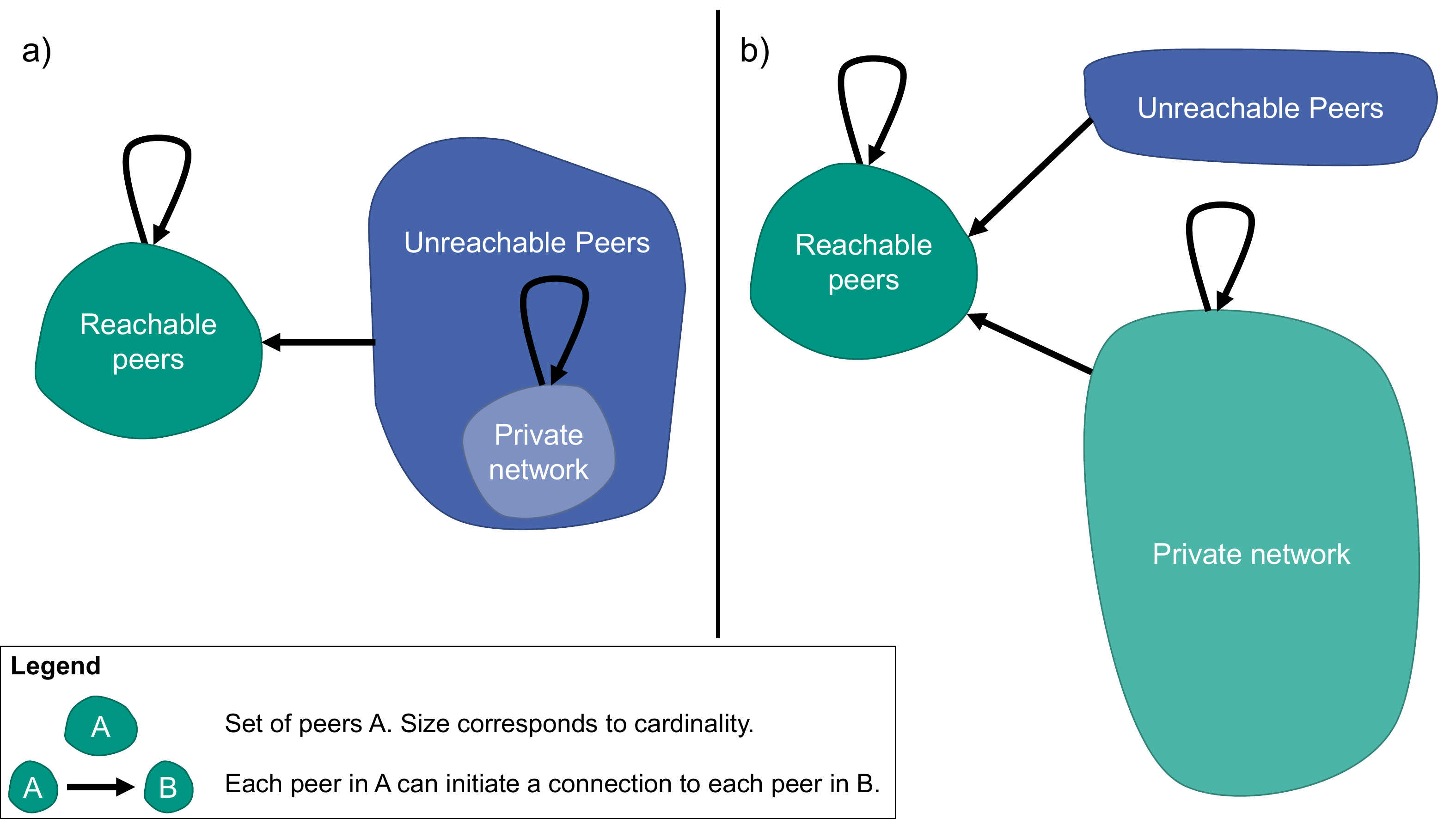}
	\caption{ Two simple models of how the peers in the Bitcoin P2P network might be connected.
				a) Each unreachable peer can initiate a connection to each reachable peer. A reachable peer can initiate a connection to another reachable peer. Additionally, there might be a private network of unreachable peers that can connect to each other.
				b) The number of peers in the private network might exceed the number of other peers. In this case, our definition would classify the peers in the private network as reachable peers although they are not reachable by peers outside the private network.
				This is not a problem for this work because we assume that the Bitcoin P2P network looks similar to the model a).
			 }
	\label{fig-unreachable-classification}
\end{figure}

Categorizing peers into reachable and unreachable peers is more difficult than it might seem at a first glance.
A first approach would be to define a peer as unreachable if all other peers cannot initiate a connection to that peer.
A peer might be unreachable because it runs behind a firewall that blocks incoming connections.
However, a peer might accept incoming connections from one group of peers but refuse incoming connections from other peers.
Imagine a private network that blocks incoming connections from the outside to peers inside the network but allows for incoming connections between peers that are inside the private network (see \cref{fig-unreachable-classification}a).
Because a peer in such a network would accept incoming connections from \textit{some} peers, this peer would not be called unreachable following the above definition of unreachable.
However, this peer would be unreachable for all peers outside the private network and, thus, should by intuition be categorized as unreachable.
Therefore, a definition of unreachable peers may not be too strict.
We consider such cases in the following definition that we use in this paper:
A peer $p$ is called \textit{unreachable} if the majority of other peers cannot initiate a connection to $p$.
This means that a peer is called reachable if most other peers can initiate a connection to that peer.
Note that, if the majority of peers were inside a private network that blocks connections from the outside but allows for internal connections, this definition would categorize the peers in this private network as reachable (see \cref{fig-unreachable-classification}b) because the majority of other peers is in the same private network.
Indeed, the peers in this private network might have many incoming connections and, thus, look like reachable peers.
However, as peers outside the private network cannot initiate a connection to them, the peers in the private network would seem unreachable to the rest of the world.
We think that a binary classification into reachable and unreachable peers would not be suitable for such a scenario.
For this work, however, we assume that the Bitcoin P2P network follows mostly the model as sketched in \cref{fig-unreachable-classification}a, i.e. most peers in the Bitcoin P2P network accept incoming connections from either all other peers or none.

Our goal is to estimate the number of unreachable peers.
Because peers join and leave the network (\textit{churn}) this number changes continuously.
The number of peers at a given point in time can be different from the number of peers that existed during a given time period (e.g., during one day).
In the following, we will talk about estimating the number of unreachable peers during time periods.
Using a model for churn, this number can be used to estimate how many unreachable peers existed at a given point in time.

The peers in the Bitcoin P2P network are identified by their addresses.
A peer can have multiple addresses (in the most common case an IPv4 address and an IPv6 address) and multiple peers can share an address (e.g., an IPv4 address because they are behind the same NAT).
In this work, we will make the simplifying assumption that each peer has exactly one address.
If we simply use the term address, then it refers to any type of address being used in the Bitcoin protocol, e.g., IPv4 address, IPv6 address, or Tor address (see BIP 155\footnote{\url{https://github.com/bitcoin/bips/blob/master/bip-0155.mediawiki}} for full list).

\section{Related Work}
\label{sec-related-work}

The number of reachable peers is continuously measured by different projects (see \cref{footnote-projects} on page \pageref{footnote-projects}).
They share the basic approach of recursively searching the network for peers.
Exemplary, we explain the approach of Bitnodes\footnote{\url{https://github.com/ayeowch/bitnodes/}} which is similar to that of \cite{park_nodes_2019-2}:
The software starts with an initial set of peers, connects to each peer and requests addresses from this peer using a \textsc{getaddr} message.
On receiving an \textsc{addr} message as reply, the software tries to connect to each of the addresses in the reply and, for each successfully opened connection, addresses are requested over this new connection.
The set of peers that a connection has been established to, is regarded as the set of reachable peers.
In case a connection to an address cannot be established, it is unknown whether there is no peer at this address or there is an unreachable peer at this address.
Therefore, this approach is not capable of measuring the number of unreachable peers.

In previous work, only few attempts have been made to estimate the number of unreachable peers.
In May 2017, Wang and Pustogarov \cite{wang_towards_2017} ran 102  reachable peers as probes in different data centers around the globe for seven days and logged all incoming connections and associated information.
For each peer that connected to one of the probes, they tried to open a TCP connection on port 8333 to that peer's IP address and to open a connection via the Bitcoin protocol.
They observed on average about 10,000 unique IP addresses in a 6-hours interval.
They assume an average of 3.5 connections per unreachable peer and 5,540 reachable peers and use these numbers to estimate that there were at least 155,000 unreachable peers in each 6-hours interval.
The authors measured that 34.6\,\% of the observed connections lasted shorter than one second and 93.9\,\% lasted shorter than one minute.

To parameterize their simulation, Naumenko et al. \cite{naumenko_erlay_2019-1} used numbers obtained from a website run by Luke-Jr\footnote{\url{https://luke.dashjr.org/programs/bitcoin/files/charts/historical.html}}.
They obtained the information that the network had 54,000 unreachable peers (called non-listening) and 6,000 reachable peers (called listening).
At the time of writing (January 2021), the website lists about 50,000 unreachable peers and 5,000 reachable peers.
The methodology behind the website is not publicly documented, but, in the absence of other reference points, we also compare our measurements to the numbers obtained from this website.

\section{Bitcoin Peers}
\label{sec-background}

The protocol for peers in the Bitcoin P2P network is described by a public developer reference\footnote{\url{https://developer.bitcoin.org/reference/index.html}}.
The protocol does not distinguish between reachable and unreachable peers because a peer cannot detect whether it is unreachable or is reachable but does not have any incoming connections, yet.
Because most of the peers run the same software, Bitcoin Core (announced in \textsc{version} messages as "Satoshi"),\footnote{\url{https://dsn.kastel.kit.edu/bitcoin/}} the behavior of this software is the de facto specification of the protocol.
In the following, we start by describing the protocol rules that all peers should follow and then describe the relevant behavior of Bitcoin Core in more detail.

\subsection{Bitcoin Protocol}

\begin{table}[tb]
    \caption{Messages of the Bitcoin protocol relevant for our study}
    \begin{center}
        \begin{tabular}{@{}*{2}{l}@{}}\toprule
        	Type & Relevant Fields\\\midrule
        	 \textsc{version} & protocol version, services, user agent, address of receiving peer, \dots \\
        	 \textsc{verack} & -- \\
        	 \textsc{getaddr} & -- \\
        	 \textsc{addr} & number of entries, for each entry: timestamp, services, address, port \\\bottomrule
        \end{tabular}
    \end{center}
    \label{tbl-protocol-messages}
\end{table}

\Cref{tbl-protocol-messages} gives an overview of the messages in the Bitcoin protocol relevant for this paper.
We briefly explain these messages in this paragraph.
Peers need to know the addresses of other peers to be able to connect to them.
To this end, addresses are propagated in the Bitcoin P2P network using \textsc{addr} messages.
An \textsc{addr} message consists of a header, the number of entries that follow, and one or multiple entries.
Each entry consists of an address, a port, a timestamp, and service flags.
The timestamp was originally meant to describe when that peer was seen last, however, the behavior of the reference software was modified to not leak which peers are currently connected \cite{miller_discovering_2015}.
The service flags describe the services offered and extensions implemented by the peer running at the address.
The protocol allows \textsc{addr} messages to contain up to 1000 entries.
\textsc{addr} messages can be sent unsolicited and they can be requested using a \textsc{getaddr} message.
A peer replies to a \textsc{getaddr} message with an \textsc{addr} message.

When a connection between two peers is established, they send \textsc{version} messages to each other that contain the peers' user agents and services.
The receiving peer replies to a \textsc{version} message by sending a \textsc{verack} message.

\subsection{Bitcoin Reference Software}

By default, Bitcoin Core opens eight outgoing connections for full-relay and two outgoing connections that are used only for relaying blocks but not for transactions and addresses.
Thus, an unreachable peer maintains outgoing connections to ten other peers.
A reachable peer can additionally have incoming connections.
The default maximum number of all (outgoing and incoming) connections is 125.
In case a new incoming connection fills the last available slot, Bitcoin Core will evict an existing connection based on different metrics such as duration of the connection, ping times, and amount of transmitted data.

Peers request addresses from other peers using the \textsc{getaddr} message which is answered by an \textsc{addr} message.
The reply contains at most 23\,\% of the addresses in the replying peers database and at most 1000 addresses.
\textsc{addr} messages are also sent unsolicitedly:
A peer announces its address to a connected peer once a connection has been established.
Each peer also regularly announces its address to its connected peers (except those for block-relay only).
The announcements are sent at random times following an exponential distribution with a mean of 24 hours.
In contrast to replies to a \textsc{getaddr} message, these announcements are forwarded so that peers in the network learn about other peers.
To distinguish between forwarded announcements and an announcement of a peer's own address, we call the later a \textit{self announcement}.

To tell whether an \textsc{addr} message was received unsolicitedly or in reply to a \textsc{getaddr} message, Bitcoin Core checks the number of entries in an \textsc{addr} message.
Because announcements of addresses are unsolicited, most of the messages with announcements are small messages with ten or less entries.
Thus, Bitcoin Core considers the addresses in an \textsc{addr} message for propagation if the \textsc{addr} message contains ten or less entries.
An address is only propagated if it meets the following requirements:
(1) The service flags associated with the address need to have the \textsc{node_witness} flag set and the \textsc{node_network} or \textsc{node_network_limited} flags.
(2) The timestamp associated with the address must not be older than ten minutes.
(3) The address itself must be routable, i.e., it may not be from an IP address range that is reserved for private use.
If the decision is for an address to be propagated, then it is sent to one or two connected peers.

\section{\methodName{} Method and Results}
\label{sec-method}

In this section, we present the \methodName{} method and give details on the setup and data collection and the methodology for analyzing the data.
We discuss the limitations of this approach and present the results of applying this method to data collected during a timespan of five years.

For the \methodName{} method, we run a passive monitor node that is connected to all reachable peers.
The monitor collects unsolicitedly sent \textsc{addr} messages that are received from its peers.
The set of all addresses received throughout a day results in an estimate of the active peers during this day.
After removing the addresses of reachable peers from this set, we have an estimate of the set of unreachable peers.

The \methodName{} method can be seen as a refinement 
on the approach for finding reachable peers explained at the beginning of \cref{sec-related-work}.
This approach itself is unsuitable for reliably finding unreachable peers because it only reveals that there are reachable peers at addresses at which an incoming connection is accepted.
If no incoming connection is accepted at an address, the approach does not allow to draw any conclusions on whether there is no peer or an unreachable peer at this address.
Instead of relying on \textsc{getaddr} messages to quickly collect many addresses, we use a passive monitor node that waits for unsolicited \textsc{addr} messages.
An address is only sent unsolicitedly if it is forwarded or if the sending peer announces this address as its own address (self announcement).
If we receive an address in an unsolicited \textsc{addr} message at the monitor, we can conclude that at most ten minutes ago there was a peer at this address because these messages are only propagated until the timestamp associated with the address is older than ten minutes.
Collecting all unsolicitedly sent addresses during one day gives us an estimate of the set of peers during this day.
By filtering out reachable peers, we receive an estimate of the set of unreachable peers for this day.

\begin{figure}[tbp]
	\centering
	\includegraphics[width=10cm]{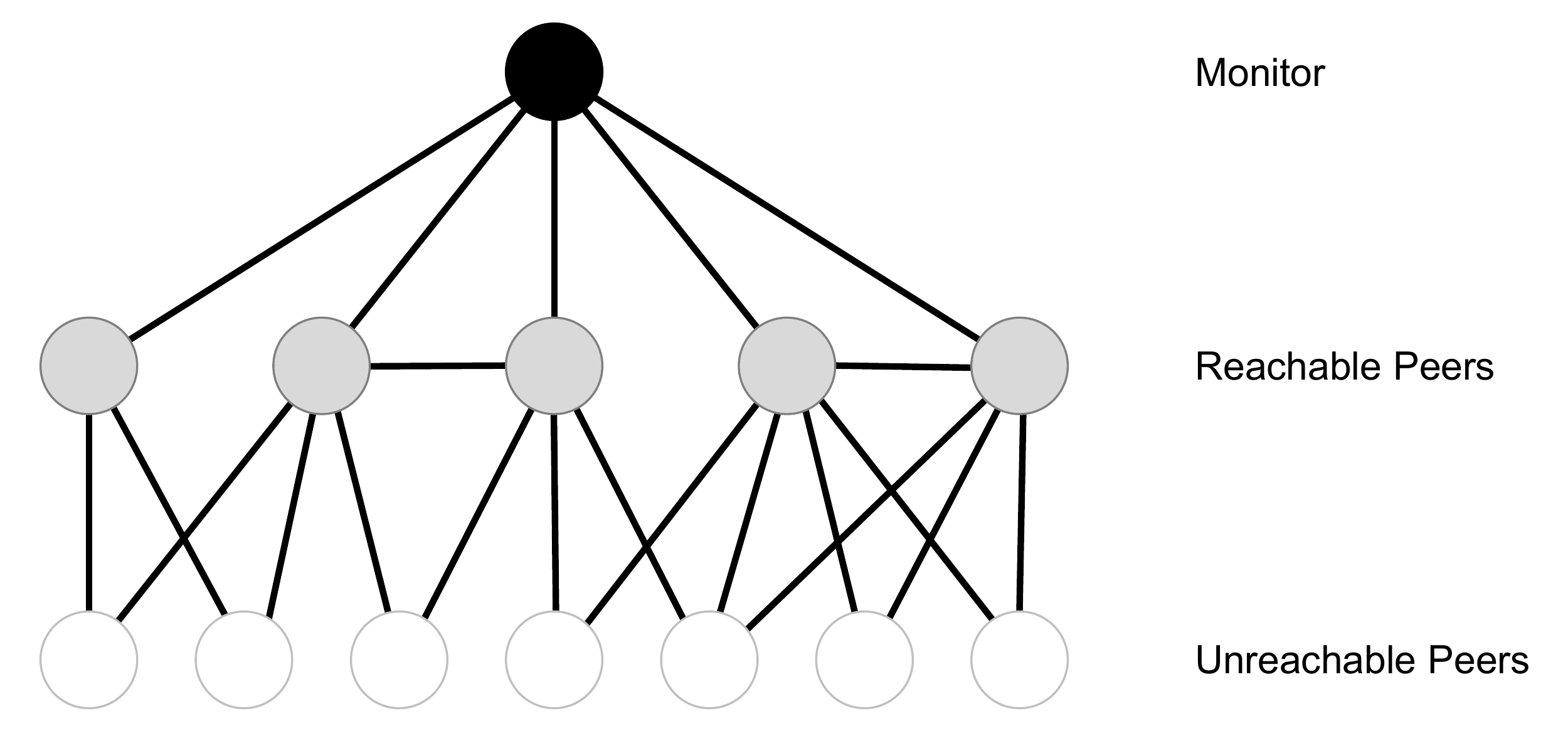}
	\caption{Overview of the setup. The monitor node is connected to all reachable peers. Unreachable peers are connected to reachable peers, too.
			 There are connections between reachable peers but no connections between unreachable peers. 
			 }
	\label{fig-setup-overview}
\end{figure}

\paragraph{Data Collection}
We run a monitor node that connects to all known reachable peers in the network (see \cref{fig-setup-overview}).
The monitor is mostly passive, i.e. it does not send any messages with the following exceptions:
\begin{itemize}
	\item After opening a connection, the monitor sends its own identifier in a \textsc{version} message
	\item After having received another peer's \textsc{version} message, the monitor not only sends a \textsc{verack} message but also requests addresses by sending a \textsc{getaddr} message.
	\item Every two minutes, the monitor sends a \textsc{getaddr} message to a uniform randomly chosen peer.
\end{itemize}
Sending the \textsc{getaddr} messages is not required for the \methodName{} method but also does not interfere with our results; the \textsc{addr} messages that are received in reply to these messages will be ignored in the following.
The monitor tries to connect to each address it receives in \textsc{addr} messages (rate-limited per address to once every six hours).
The monitor logs all received \textsc{addr} messages, \textsc{version} messages and the time when a connection to another peer is established or closed.
    
\begin{table}[bt]
    \caption{Overview of notation (also see \cref{fig-data-analysis})}
    \vspace*{-4mm}
    \begin{center}
        \begin{tabular}{@{}*{2}{l}@{}}\toprule
        	Symbol & Description\\\midrule
        	 $\addrIPs{t}$ & Set of addresses received on day $t$ in small \textsc{addr} messages (senders excluded) \\
        	 $\publicIPs{t}$ & Set of addresses that the monitor node was connected to on day $t$ \\
        	 $\estimateUnreachable{t}$ & (Estimation of) Set of addresses of unreachable peers on day $t$ \\
        \end{tabular}
    \end{center}
    \label{tbl-notation}
    
    \vspace*{3mm}
    
    \centering
	\includegraphics[width=\textwidth]{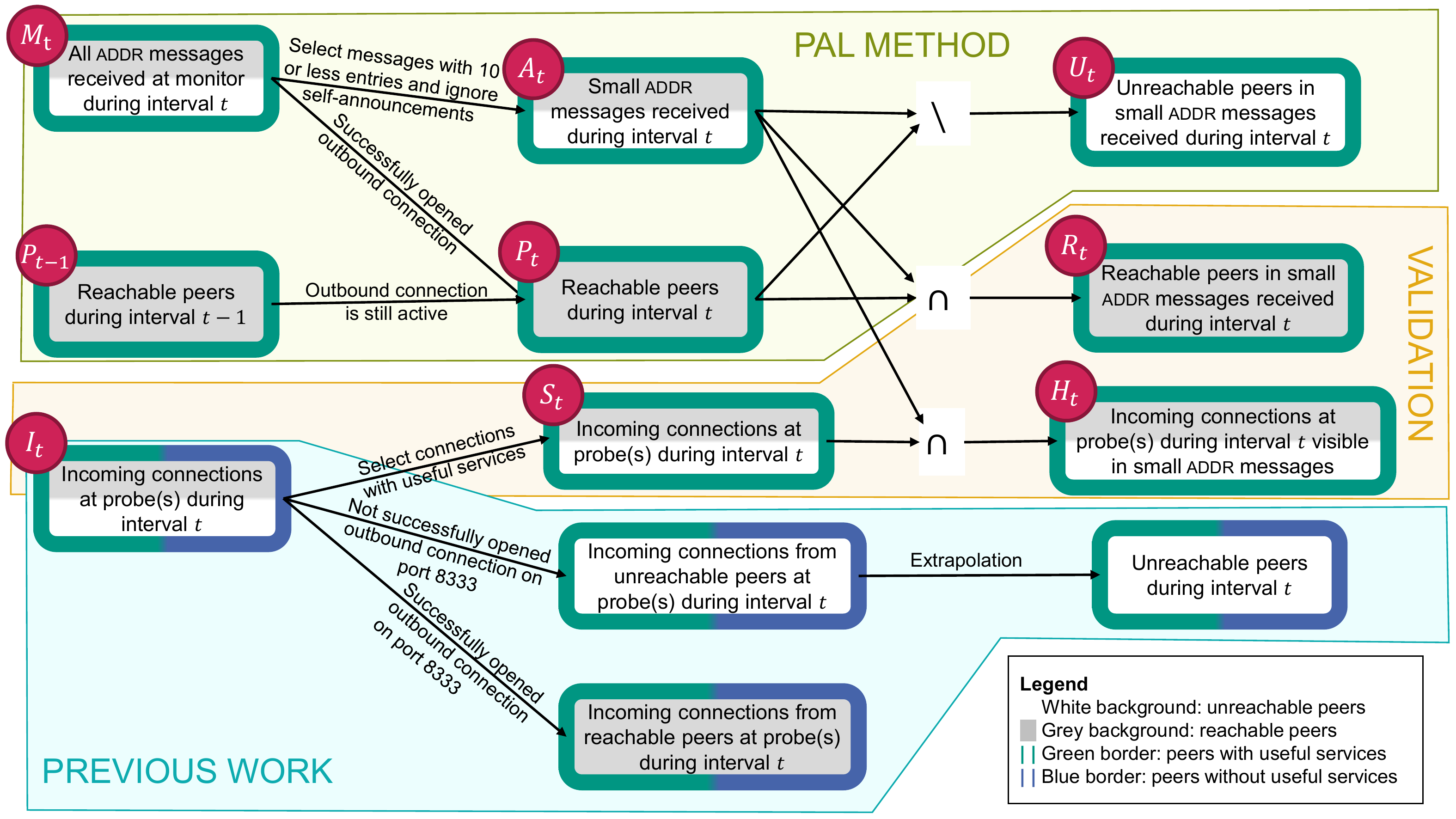}
	\vspace*{0mm}
	\captionof{figure}{Data flow of the \methodName{} method and the approach of \cite{wang_towards_2017} to estimate the number of reachable and unreachable peers in the Bitcoin P2P network.
	The sets $M_t$ and $I_t$ are collected during measurements and the arrows show filters and operations to derive more specific sets during the analysis.
	The border color of each box indicates whether the respective set contains peers with useful services only or also those without.
	The background color of each box indicates whether the respective set includes reachable and unreachable peers.
			 }
	\label{fig-data-analysis}
\end{table}

\paragraph{Data Analysis}
We analyze the logs created by the monitor with the goal of learning the number of peers in the network.
We describe this process in the following and depict it in the upper part of \cref{fig-data-analysis}.
The most relevant notation symbols are listed in \cref{tbl-notation}.
For each day $t$, we collect all addresses that were received by the monitor (see $M_t$ in \cref{fig-data-analysis}).
We define the set $\addrIPs{t}$ by applying the following two filters on these addresses:
(1) We select only the addresses in small \textsc{addr} messages (we define \textit{small} for \textsc{addr} messages as containing ten or less entries).
(2) We ignore the self announcements of (reachable) peers, i.e. entries of an \textsc{addr} message that equal the address of the sender of this \textsc{addr} message.
The set $\addrIPs{t}$ includes addresses of reachable and unreachable peers that were announced on this day.
We determine the set $\publicIPs{t}$ of all addresses that the monitor node was connected to on day $t$.
For this, we collect all addresses that the monitor already was connected to at the beginning of day $t$ or a connection was established and a \textsc{version} message received during day $t$.
We consider this set $\publicIPs{t}$ as the set of all reachable peers at day $t$.
Our estimate of the unreachable peers $\estimateUnreachable{t}$ for day $t$ is $\estimateUnreachable{t} = \addrIPs{t} \setminus \publicIPs{t}$.

\paragraph{Limitations}
As described in \cref{sec-background}, peers running Bitcoin Core only forward addresses that have specific flags set (\textsc{node_witness} and (\textsc{node_network} or \textsc{node_network_limited})).
Thus, we do not expect to receive addresses of peers that do not have these flags set at the monitor node.
The \methodName{} method can only detect such peers with useful services which is indicated by the border colors in \cref{fig-data-analysis}.

Another limitation of the \methodName{} method is that it cannot distinguish whether an unreachable peer existed only for a short moment on a day or the whole day.
Also, the addresses and associated information in \textsc{addr} messages are not authenticated.
Thus, a peer can send any address and information about this address to other peers in the network.
If a peer sends addresses in an \textsc{addr} message, there is no proof that the peer has received the address from another peer or is connected to a peer with this address.
Therefore, the approach can be disturbed by flooding the network with bogus addresses.

\paragraph{Measurements}

\begin{figure}[tbp]
	\centering
	\includegraphics[width=15cm]{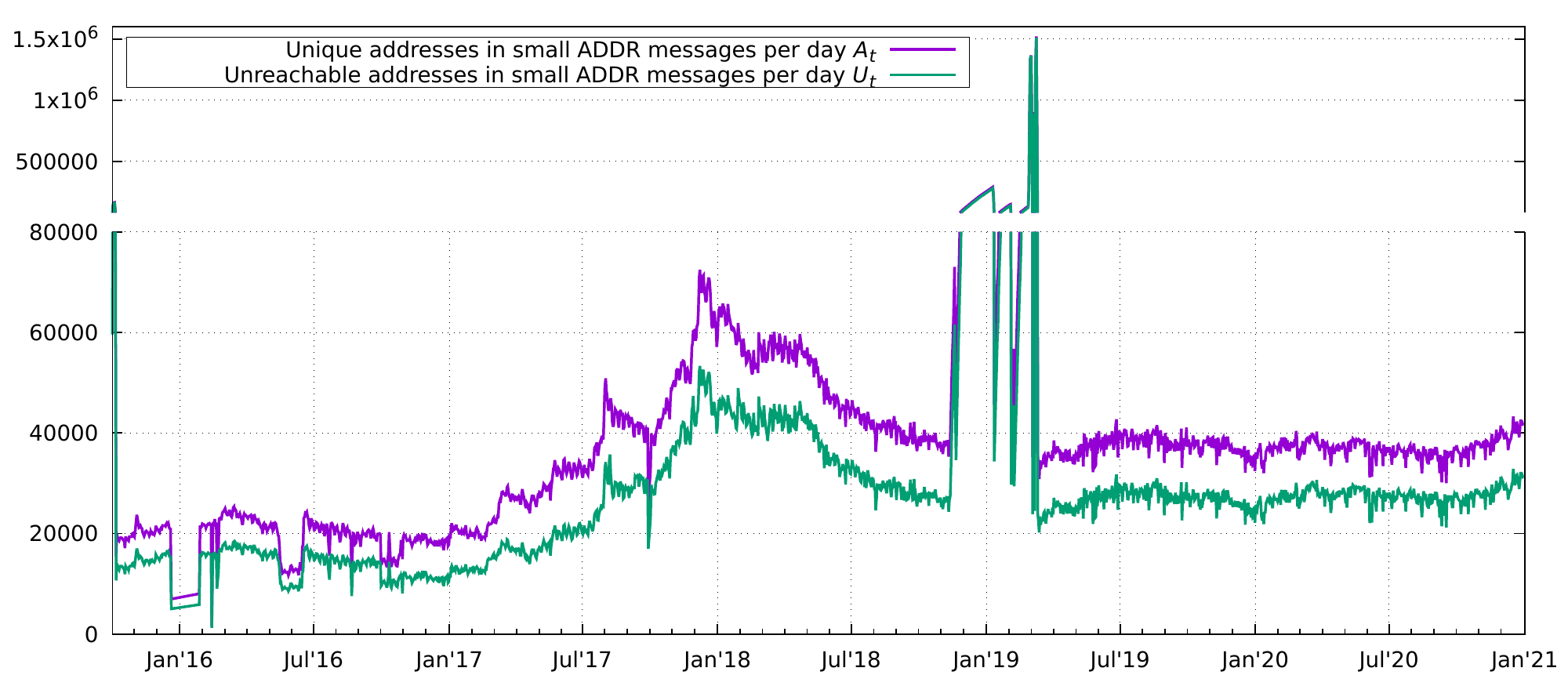}
	\caption{Number of unique addresses observed in small \textsc{addr} messages per day. Note that the upper part uses a different scale than the lower part.}
	\label{fig-unique-small-day}
\end{figure}

We applied the method to data collected from 2015 to 2020 and present the results here.
\Cref{fig-unique-small-day} shows $\vert \addrIPs{t} \vert$, the number of unique addresses received in small messages for each day $t$ and the number $\vert \estimateUnreachable{t} \vert$ of addresses that were unreachable.
The plot shows that, at the majority of days in the observation range, between 20,000 and 60,000 unique addresses were received in small messages.
Most noticeably, the plot shows a high amount of addresses at the end of the year 2018 and beginning of the year 2019 which we will discuss later.
The remaining plot shows that the number of addresses increased from the beginning of the year 2017 on to a maximum of about 72,000 addresses in the middle of December 2017.
From this point on, the number of addresses declined slowly until the prominent peaks around the turn of the year 2018/2019 and since then the number of addresses has been relatively stable around 37,000.
The number of unreachable peers $\vert \estimateUnreachable{t} \vert$ has a similar development but is consistently about 28\,\% lower than the number of all addresses.
At the end of the year 2020, the number of unreachable peers $\vert \estimateUnreachable{t} \vert$ equals about 31,000 peers.

The peak at the end of the year 2018 seems like many unreachable peers joined the network within a few days.
An alternative explanation would be that bogus addresses were distributed that do not actually belong to peers.
We examined the addresses that were received only during this time and did not find any irregularities with regard to their distribution in the IP address space, autonomous system, or country of autonomous system.
However, for the highest peak in March 2019, we found that this peak was caused by many IP addresses from the same /8 subnet.
As IP addresses from this subnet were only very rarely observed before and after March 2019, we assume that this effect was caused by unknown actions of one party that flooded the network with these IP addresses.
It is an open question whether a known or unknown attack would cause such effects and whether this can be verified using the data from our measurements.

\section{Validation}
\label{sec-validation}

To validate the \methodName{} method, in this section we analyze the measured data and compare it to other sources.

\subsection{Reachable Peers}

As a first step, we verify that the results of the \methodName{} method are consistent in itself.
If our assumptions hold true, we should be able to find all reachable peers in small \textsc{addr} messages during each day.
As we know the set of reachable peers, we can use it to evaluate the precision of our measurements.
Putting this into the context of \cref{fig-data-analysis}, this means that, if the \methodName{} method works perfectly, we expect that set $R_t$ equals set $P_t$.

\begin{figure}[tbp]
	\centering
	\includegraphics[width=15cm]{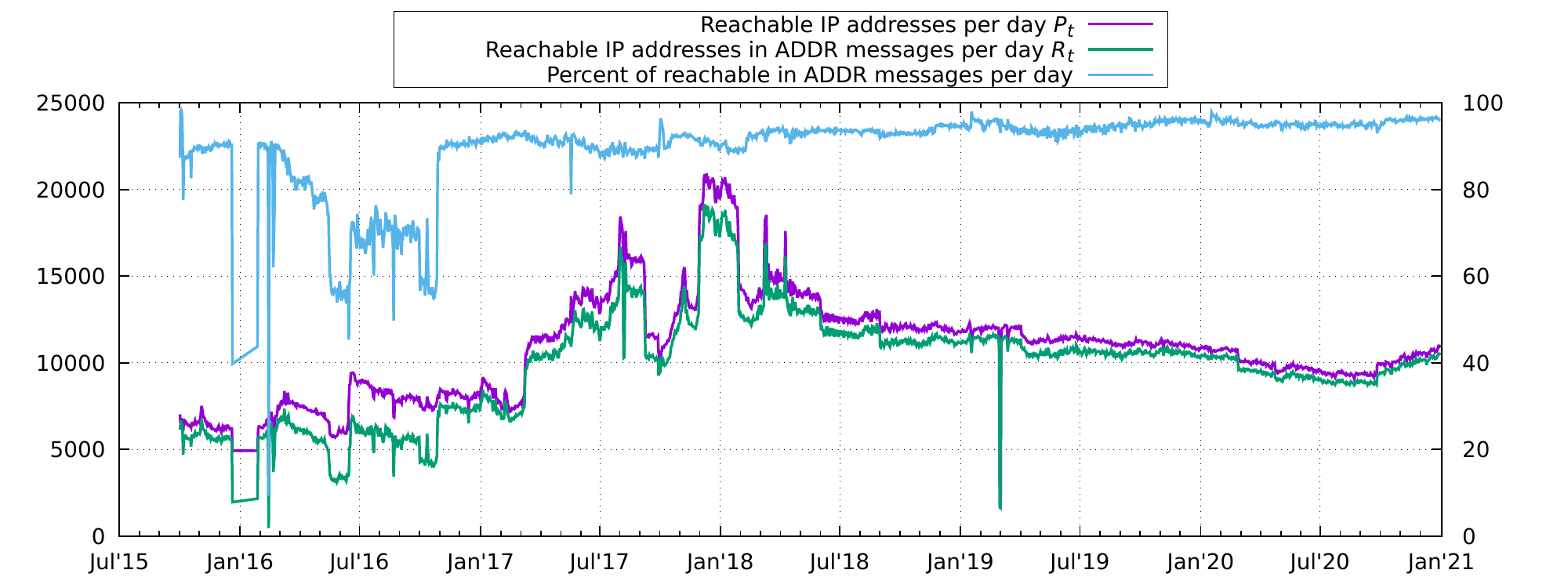}
	\caption{Number of unique addresses of reachable peers observed in small \textsc{addr} messages per day.}
	\label{fig-public-in-addr-per-day}
\end{figure}

In \cref{fig-public-in-addr-per-day}, we plot the number $\vert \publicIPs{t} \vert$ of reachable peers for each day $t$ and the number $\vert \addrIPs{t} \cap \publicIPs{t} \vert = \vert R_t \vert$ of reachable peers whose addresses we received on each day $t$.
The plot shows that since July 01, 2016 on each day on average 91.4\,\% of the addresses of reachable peers were received in a small \textsc{addr} message on the same day (excluding events for peers sending their own address).
Over the time, the quality has increased: Since Jan 01, 2017 the average is 93.4\,\% and since Jan 01, 2020 the average is 95.3\,\%.
This is a promising result as it indicates that reachable peers are consistently found by the \methodName{} method.
However, as can be seen in \cref{fig-data-analysis}, we do not have an independent ground-truth for the set of reachable peers.
Because all these sets of peers are based on the contents of \textsc{addr} messages, we can only conclude that the \methodName{} method's constraint of looking at one day only and at small messages only does not reduce the set of detected reachable peers.

\subsection{Second Monitor}

\begin{figure}[tbp]
	\centering
	\includegraphics[width=15cm]{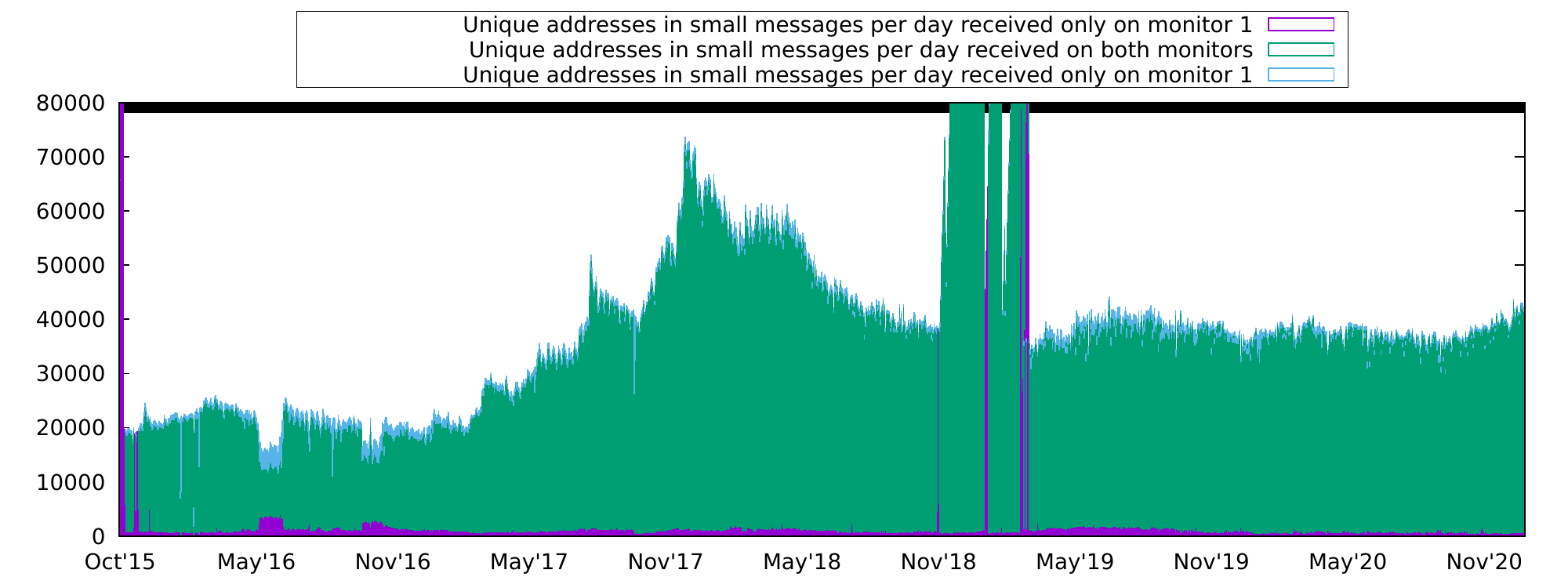}
	\caption{Stacked plot of the number of unique addresses observed in small \textsc{addr} messages per day at two monitors. The plot is separated into the number of addresses received only at monitor 1 (purple), the number of addresses received at both monitors (green), and the number of addresses received only at monitor 2 (blue). It can be seen that the majority of addresses is received at both monitors.}
	\label{fig-common-in-monitors-per-day-stacked}
\end{figure}

\begin{figure}[tbp]
	\centering
	\includegraphics[width=15cm]{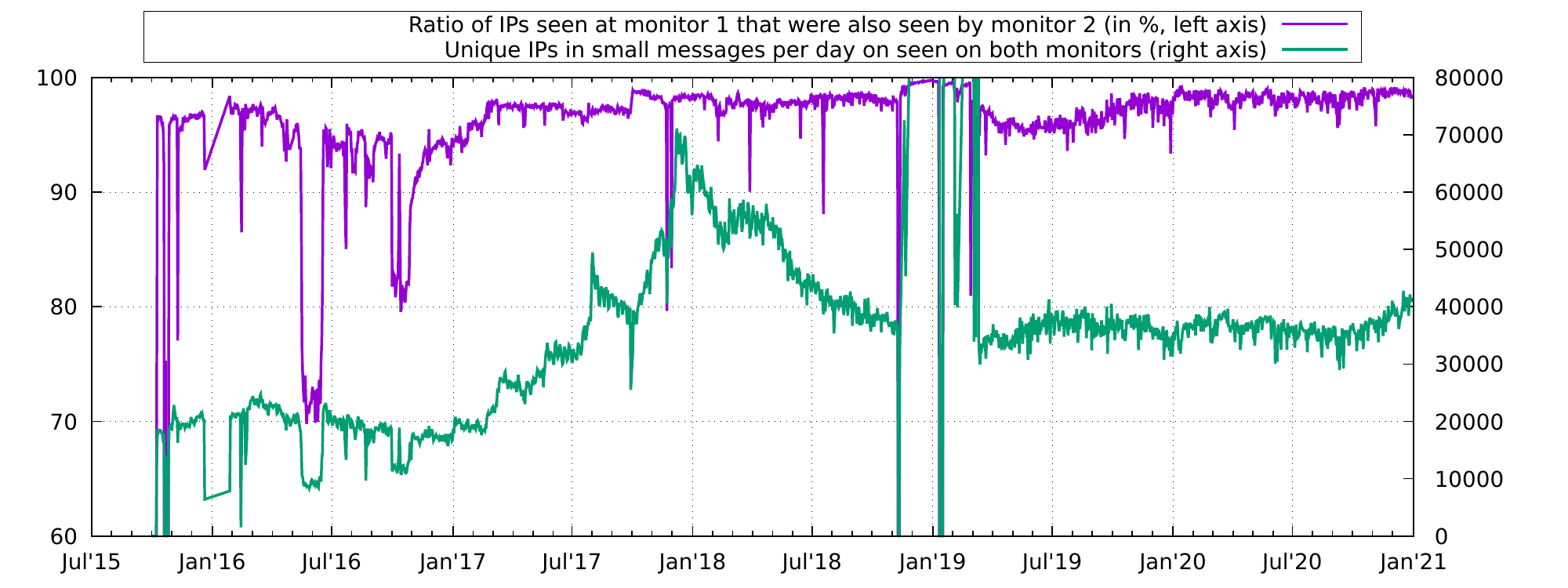}
	\caption{Ratio of IPs seen at monitor 1 that were also seen at monitor 2. The green plot shows the number of unique addresses observed in small \textsc{addr} messages per day at both monitors (right y-axis). The green plot is the same as the green plot in \cref{fig-common-in-monitors-per-day-stacked}.}
	\label{fig-common-in-monitors-per-day}
\end{figure}

For further validation, we run a second monitor node with the same method as describe above.
We compare the addresses received by two monitor nodes.
If the measurement method is reproducible, the data of both monitors should largely overlap.
\Cref{fig-common-in-monitors-per-day-stacked} and \cref{fig-common-in-monitors-per-day} show the results for two monitor nodes for the whole time of measurement.
It can be seen that since the beginning of 2017, over 95\,\% of the addresses observed at monitor 1 are also observed at monitor 2.
This indicates that the measurement is reproducible and that the view of our monitor node is not subjective to the specific instance of the monitor.

\subsection{Unreachable Peer}

If our assumptions hold true, we can expect an unreachable peer that runs continuously to be visible in small \textsc{addr} messages every day.
We run a small experiment to test this hypothesis.
We start an unreachable peer $p_\mathrm{U}$ running the Bitcoin Core software in version v0.20.1 with the only modification to create a log entry when the peer announces its own address.
The peer $p_\mathrm{U}$ has an IPv4 and an IPv6 address.
The log entry consists of a timestamp, which address is announced, and which peer the address is announced to.
The peer $p_\mathrm{U}$ runs continuously and it is unreachable because its incoming TCP port 8333 is blocked by a firewall.
After being started, $p_\mathrm{U}$ creates ten outgoing connections (eight for full-relay and two for block-relay only).
After the peer has been continuously running for $44$ days, we analyze the logs and compare them to our monitoring logs.
We first look at the number of announcements that the peer $p_\mathrm{U}$ has sent on each day.
While one might expect that $p_\mathrm{U}$ sends on average eight announcements per day because it announces its address to every connected full-relay peer on average every 24 hours, the average measured is about 16. 
This is because although the peer $p_\mathrm{U}$ is running continuously, connections are closed by other peers and $p_\mathrm{U}$ creates new outgoing connections on which $p_\mathrm{U}$ announces its address.
The number of announcements sent on a day is composed of the announcements on newly created connections and regular announcements on existing connections.
This fact increases our expectations to receive an address of $p_\mathrm{U}$ at the monitor on every day.
Next, we examine this proposition by looking for $p_\mathrm{U}$'s addresses in the monitor's logs.
We find that the monitor received $p_\mathrm{U}$'s address on 43 of the 44 days.
Assuming that this observation can be generalized, this indicates that the monitor sees the addresses of continuously running peers running the Bitcoin Core software with high probability on each day and the \methodName{} method can detect these peers.
We leave validation of this hypothesis for other Bitcoin software for future work.

\subsection{Validation with Incoming Connections at a Public Peer}
\label{sec-validation-incoming}

The approach of \cite{wang_towards_2017} is to run many public peers that accept incoming connections and observe the connections they receive which are partially from unreachable peers.
For further validation of the \methodName{} method, we use a similar approach and run a single public peer $p_\mathrm{I}$ that accepts incoming connections.
We compare the collected data to the data found in \textsc{addr} messages to find out how many unreachable peers seen by $p_\mathrm{I}$ are also found using the \methodName{} method.

The peer $p_\mathrm{I}$ does not open outgoing connections\footnote{Before the experiment, a peer existed that had the same address as $p_\mathrm{I}$ and initiated outgoing connections. Therefore, $p_\mathrm{I}$'s address is contained in other peer's databases and peers open incoming connections to $p_\mathrm{I}$.} and logs for each connection when it is established and closed and the received \textsc{version} messages.
We analyze the logs after $p_\mathrm{I}$ has been running for about 16 months.
On average, the peer $p_\mathrm{I}$ had incoming connections from about 2,040 different addresses per day (this corresponds to the size of $I_t$ in \cref{fig-data-analysis}).
As \textsc{addr} messages contain almost only addresses of peers providing useful services, we only select such addresses for comparison.
The resulting set per day contains on average about 946 addresses (this corresponds to $\vert S_t \vert$ in \cref{fig-data-analysis}).
The intersection of this set with all addresses received in small \textsc{addr} messages on the same day, results in the set of addresses that opened a connection to $p_\mathrm{I}$, announced useful services, and were observed in a small \textsc{addr} on the same day (see $H_t$ in \cref{fig-data-analysis}).
If the \methodName{} method worked perfectly, than this set $H_t$ would equal the set $S_t$ in \cref{fig-data-analysis}.
We find that on average 68\,\% of the addresses of incoming connections with useful services were also observed in small \textsc{addr} messages on the same day.
As can be seen in \cref{fig-data-analysis}, the set $H_t$ contains reachable and unreachable peers.
To validate the detection of unreachable peers only, we reduce this set to unreachable peers only by subtracting the set of reachable peers $P_t$.
This shows that about 51\,\% of incoming connections of unreachable peers are detected by the \methodName{} method in small \textsc{addr} messages.
Analogously, we find that the \methodName{} method detects about 98\,\% of incoming connections of reachable peers.

These results show that the \methodName{} method is very reliable to detect reachable peers and it detects about half of the unreachable peers with useful services.
It is an open question why unreachable peers are not detected as reliably as reachable peers.
Possible explanations could be that unreachable peers exist in the network for a shorter time than reachable peers, that unreachable peers use different software with a different behavior, or that they propagate an address that is different from the address they use for connections.

\subsection{Comparison to Previous Measurements}

As there is no ground truth that we could compare the data from the \methodName{} method to, we compare it to measurements created by previous works.
In 2014, Biryukov et al. \cite{biryukov_deanonymisation_2014} estimated that the Bitcoin P2P network had a size of 100,000 peers of which 90\,\% were estimated to be unreachable.
This estimation is too early to be compared to our data but, taking the unreachable peers without useful services into account, this estimation could fit in with our observations.
Neudecker et al. \cite{neudecker_timing_2016} simulated the Bitcoin P2P network in 2016 and estimated from the simulated propagation behavior that the P2P network had about 16,000 unreachable peers that participate in the propagation of transactions and blocks.
The \methodName{} method calculates about 14,000 unreachable peers per day averaged over the year 2016.
As the results of \cite{neudecker_timing_2016} are given for one point in time and the \methodName{} method estimates the number of unreachable peers during one day, we would rather expect that the \methodName{} method would find more unreachable peers than \cite{neudecker_timing_2016}.
Our explanation is that the lower number of unreachable peers detected might again be caused by peers not announcing useful services or by the effects mentioned at the end of \cref{sec-validation-incoming}.

A measurement of unreachable peers was conducted by Wang and Pustogarov in 2017 \cite{wang_towards_2017}.
They ran more than 100 reachable peers to get incoming connections.
From their measurements, they estimated at least 155,000 unreachable peers to be active in each 6-hours time period.
This estimate is higher than the results obtained through the \methodName{} method.
An explanation for this is that Wang and Pustogarov report that 80\,\% of unreachable peers were mobile peers that had only short-lived connections.
We assume that these peers either did not announce their IP addresses or that they did not provide useful services.
In this case, they would be invisible to the \methodName{} method which explains the difference to our results.

\begin{figure}[tbp]
	\centering
	\includegraphics[width=15cm]{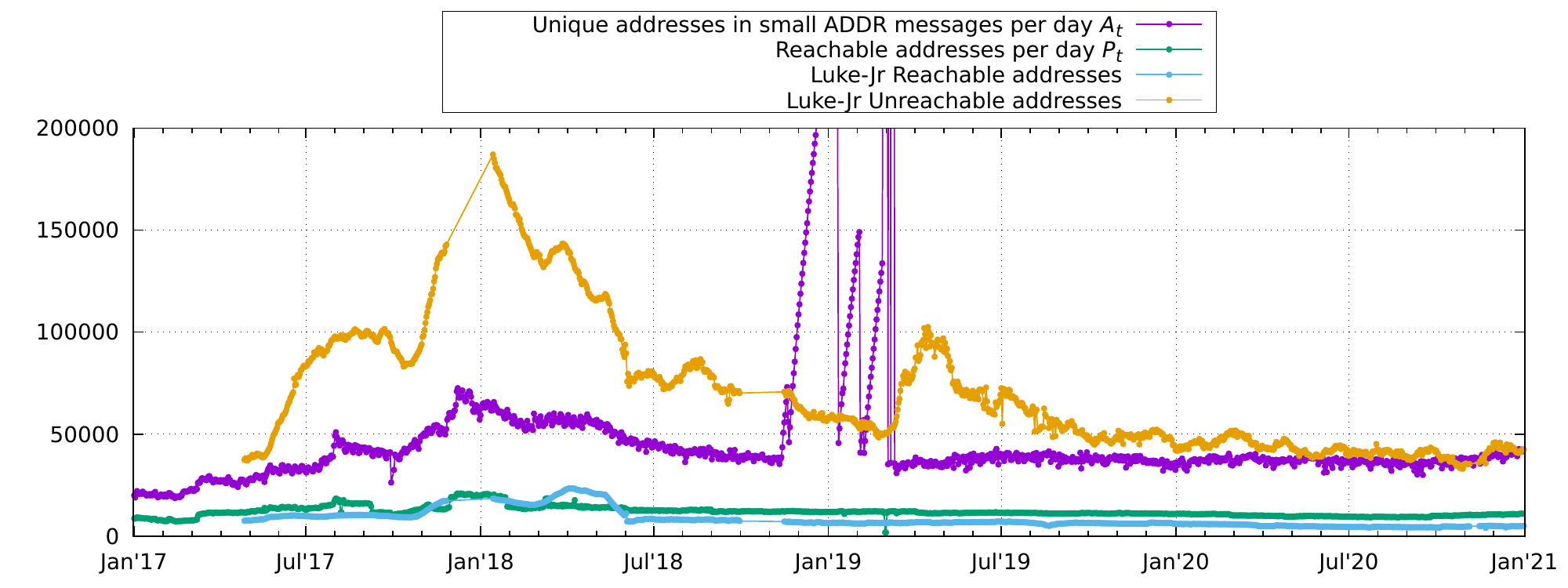}
	\caption{Comparison between the number of addresses observed in small \textsc{addr} messages per day and the data obtained from the website run by Luke-Jr.}
	\label{fig-comparison-luke-jr}
\end{figure}

The measurement by Luke-Jr\footnote{\url{https://luke.dashjr.org/programs/bitcoin/files/charts/historical.html}} gives an estimate of the number of reachable and unreachable peers over a similar timespan.
In \cref{fig-comparison-luke-jr} we plot our data and the data from Luke-Jr in one plot for comparison.
It can be seen that the number of unreachable peers in the data from Luke-Jr is higher compared to the number of addresses in small messages which is probably accounted for by the restriction that only addresses of peers providing useful services are propagated.
The increase in addresses in small messages at the end of the year 2018 cannot be seen in the data from Luke-Jr which indicates that these peaks are not caused by many peers joining the network during this time.
However, it seems that this period has had an effect on the measurements by Luke-Jr, too, because once the number of addresses in small messages had declined in March 2019, the number of unreachable peers as measured by Luke-Jr starts to rise.
One interpretation of this fact is that the high number of \textsc{addr} messages at this time, hid unreachable peers from Luke-Jr's methodology and they became visible only after the number of \textsc{addr} messages had stabilized.

\section{Conclusion}
\label{sec-conclusion}

We have presented the \methodName{} method to estimate the number of peers in the Bitcoin P2P network including unreachable peers.
The \methodName{} method adds a further reference point to existing approaches measuring the number of unreachable peers.
The estimate of the number of unreachable peers at the end of 2020 is about 31,000 peers which, as indicated by our validation, might correspond to about 50\,\% of all peers providing useful services.
Looking at the measurements of the past five years, we have seen that the number of peers varied.
As previous work considered only shorter time intervals, their measurements might have been during a time when the number of peers was higher than today.
The insights gained from this work can be valuable for development and parameterization of simulators that model the Bitcoin P2P network more realistically.

To further improve on the \methodName{} method and the validation, we plan to simulate the propagation of \textsc{addr} messages.
We expect that this can enhance our understanding of the method and of effects that are visible in the measurements.

\nocite{tange_gnu_2020}

\bibliographystyle{splncs04}
\bibliography{library}

\end{document}